\documentclass[runningheads]{llncs}
\usepackage{amssymb}
\usepackage{graphicx}
\usepackage{verbatim}
\usepackage{url}
\usepackage{enumerate}
\urldef{\mailsa}\path|xiaowen.zhang@csi.cuny.edu|
\urldef{\mailsb}\path|cchum@gc.cuny.edu|
\newcommand{\keywords}[1]{\par\addvspace\baselineskip
\noindent\keywordname\enspace\ignorespaces#1}

\begin{document}

\mainmatter

\title{Hash function based secret sharing scheme designs}

\titlerunning{Hash function based secret sharing scheme designs}

\author{Chi Sing Chum$\,^1$ and Xiaowen Zhang$\,^{1,2}$}

\authorrunning{Hash function based secret sharing scheme designs}

\institute{$^1$~Computer Science Dept., Graduate Center, CUNY\\
365 Fifth Ave., New York, NY 10016, U.S.A.\\
$^2$~Computer Science Dept., College of Staten Island, CUNY\\
2800 Victory Blvd, Staten Island, NY 10314, U.S.A.}

\maketitle

\begin{abstract}

Secret sharing schemes create an effective method to safeguard a
secret by dividing it among several participants. By using hash
functions and the herding hashes technique, we first set up a $(t+1,
n)$ threshold scheme which is perfect and ideal, and then extend it
to schemes for any general access structure. The schemes can be
further set up as proactive or verifiable if necessary. The setup
and recovery of the secret is efficient due to the fast calculation
of the hash function. The proposed scheme is flexible because of the
use of existing hash functions.

\keywords{Cryptographic hash function, herding hashes technique,
secret sharing scheme, access structure.}
\end{abstract}

\section{Introduction}

A secret sharing scheme has a strong motivation on private key
protection. Based on Kerchhoffs's principle \cite{KL07}, only the
private key in an encryption scheme is the secret and not the
encryption method itself. When we examine the problem of maintaining
sensitive information, we will consider two issues: availability and
secrecy. If only one person keeps the entire secret, then there is a
risk that the person might lose the secret or the person might not
be available when the secret is needed. On the other hand the more
people who can access the secret, the higher the chance the secret
will be leaked. A secret sharing scheme (hereafter in this paper
might be simply referred to as `scheme') is designed to solve these
issues by splitting a secret into shares and distributing these
shares among a group of participants. The secret can only be
recovered when the participants of an authorized subset join
together to combine their shares.

Secret sharing schemes have applications in the areas of security
protocols, for example,  database security and multiparty
computation (MPC).  When a client wants to have his database
outsourced (or so called ``Database as a Service'') to a third
party, how to make sensitive information hidden from the server is a
major concern. One common technique is to encrypt the data before
storing it in the server. However, queries to the encrypted database
are expensive. \cite{AAEM09} suggested to use a threshold secret
sharing scheme to split the data into different servers as shares to
handle data privacy. MPC was first introduced in Yao's seminal two
millionaires's problem \cite{YAO82}. A secure MPC can be defined as
$n$ parties $P_1, P_2, \ldots, P_n$ join together to calculate a
joint function $f(x_1, x_2, \ldots, x_n)$, where $x_i$ is the
private input by $P_i, \; i = 1, \ldots, n$. After the computation,
each $P_i$ will know the correct result of $f$ but will not know
other $x$'s. Secret sharing schemes play an important role in secure
MPC as secrecy is highly required in such computations. For more MPC
materials please refer to \cite{CDN09}.

To summarize, a secret sharing scheme is a cryptographic primitive
with many applications, such as PGP (Pretty Good Privacy) key
recovering, visual cryptography, threshold cryptography, threshold
signature, etc, in addition to those discussed above.

In this paper, we use the herding hashes technique to design a
$(t+1, n)$ threshold scheme which is perfect and ideal. Then, we
show by examples of a hierarchical threshold scheme and a
compartment scheme, that any general access structure can be
realized. The resulting scheme can be further implemented as
proactive easily. By adding an additional hash function we can make
it verifiable. The setup is simple and the secret can be recovered
quickly. The implementation is flexible as we can make use of
existing hash functions.

The rest of paper is organized as follows. In Section 2 and Section
3 we review cryptographic hash functions and secret sharing schemes.
Section 4 analyzes the complexity of the proposed scheme, and shows
how to make the implementation practical. Then, we present several
secret sharing scheme setups for illustration. In Section 5 we
outline an implementation plan. In section 6 we conclude the paper
and summarize the advantages of the proposed schemes.

\section{Cryptographic hash functions}

\subsection{Iterative hash functions and multicollisions}

A cryptographic hash function $H$ takes an input message $M$ of
arbitrarily length and outputs a fixed-length string $h$. The output
$h$ is called the hash or message digest of the message $M$. It
should be fast, preimage, second preimage and collision resistant.
Please refer to the textbooks, such as \cite{STI05,TW06}, for the
details.

An iterative hash function $H$ is basically built from iterations of
a compression function $C$ using the Merkle-Damg{\aa}rd construction
\cite{DAM89,MER79}. Briefly, the construction repeatedly applies the
compression function as follows. (a) Pad the arbitrary length
message $M$ into multiple $v$-bit blocks: $m_1, m_2, \ldots, m_b$.
(b) Iterate the compression function $h_i = C(h_{i-1}, m_i)$, where
$h_i$ and $h_{i-1}$ are intermediate hashes of $u$-bit strings,
$h_0$ is the initial value (or initial vector) IV, and $i$ ($1 \leq
i \leq b$) is an integer. (c) Output $h_b$ as the hash of the
message $M$, i.e., $H(M) = h_b = C(h_{b-1}, m_b)$.

Suppose we apply the birthday attack to get $b$ pairs of blocks
$(m_1, m_1'), \ldots, (m_{b}, m_{b}')$ such that

\begin{equation}
h_i = C(h_{i-1}, m_{i}) = C(h_{i-1}, m_{i}'), i = 1, \ldots, b.
\end{equation}

\begin{figure}
\centering
\includegraphics[height=2cm, width=8cm]{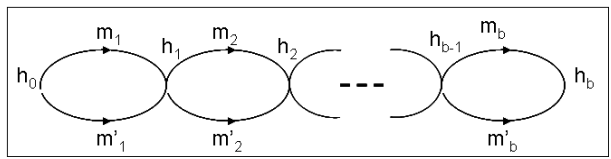}
\caption{Multicollisions in iterative hash functions.}
\label{fig:ihmulticollision}
\end{figure}

\noindent By enumerating all possible combinations of these
$b$-pairs blocks with each pair containing two choices, we can build
up $2^b$ colliding messages as follows (see
Fig.~\ref{fig:ihmulticollision}). Since it takes $2^{u/2}$ steps for
finding one pair of blocks, this process takes approximately $b
\times 2^{u/2}$ steps. So, it is relatively easy to find
multi-collisions in an iterative hash function. Please refer to
\cite{TW06,JOUX04} for the details.

\subsection{Herding and Nostradamus attack}

\begin{figure}
\centering
\includegraphics[height=5cm, width=8cm]{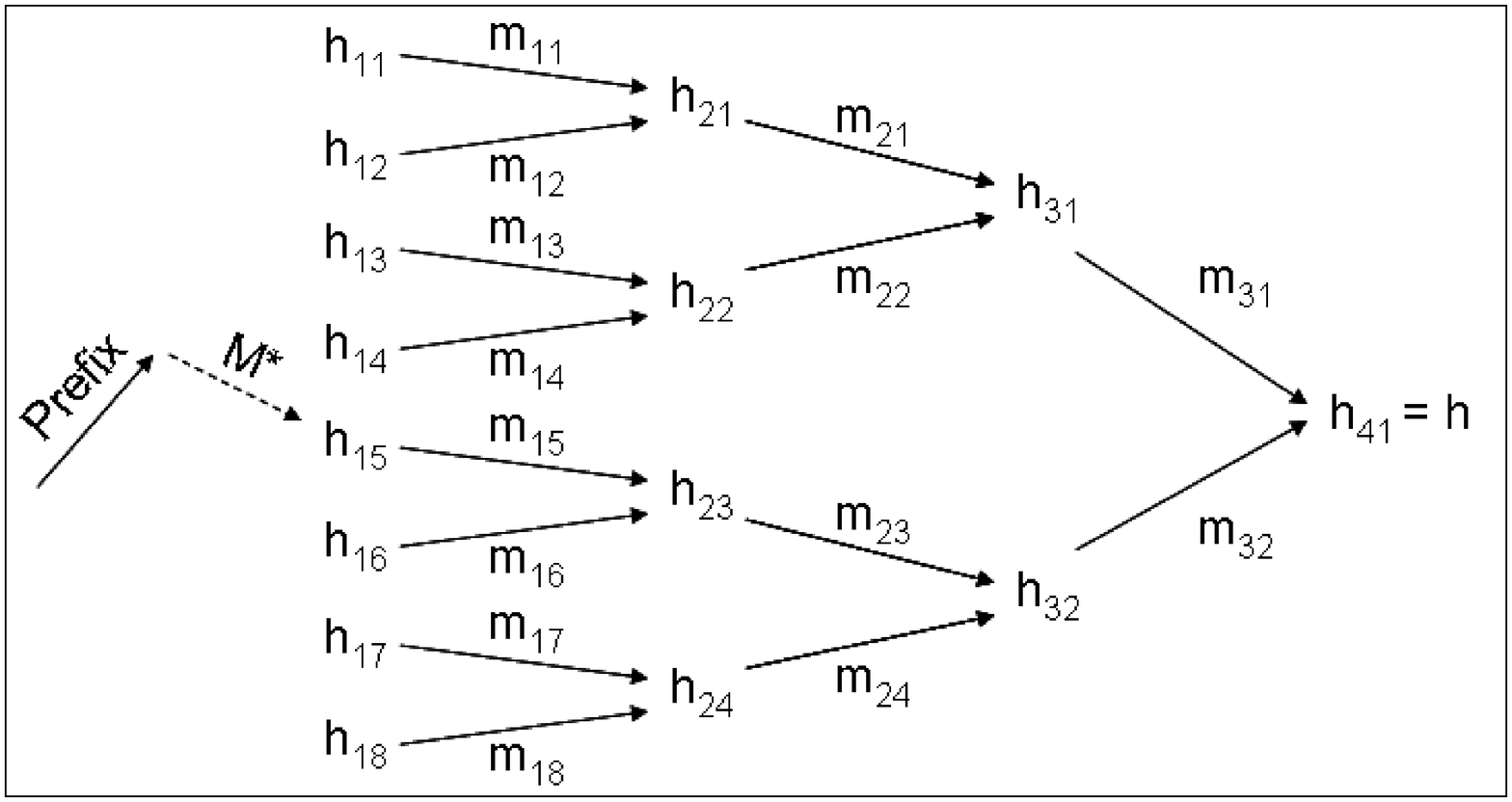}
\caption{A simplified diamond structure to illustrate Nostradamus
attack.} \label{fig:DIAM}
\end{figure}

Kelsey and Kohno \cite{KK06} have a detailed analysis of this
attack. Stevens, Lenstra and Weger \cite{SLW07} applied the
technique to predict the winner of the 2008 US Presidential
Elections using a Sony PlayStation 3 in November 2007. We first
build a large set of intermediate hashes at the first level:
$h_{11}, h_{12}, \ldots, h_{1w}$. Then message blocks are generated,
so that they are linked and each intermediate hash at level 1 can
reach the final hash, say $h$. This is called the diamond structure
(see Fig. \ref{fig:DIAM}). We claim we can predict that something
will happen in the future by announcing the final hash to the
public. When the result is available, we construct a message as
follows:

\begin{equation}
M = \mbox{Prefix} \| M^* \| \mbox{Suffix},
\end{equation}

\noindent where ``Prefix'' contains the results that we claimed we
knew before it happens. $M^*$ is a message block which links the
``Prefix'' to one of the intermediate hashes at level 1. ``Suffix''
is the rest of message blocks which linked the $M^*$ to the final
hash. In the example of Fig.~\ref{fig:DIAM}, $M = \mbox{Prefix} \|
M^* \| \mbox{Suffix}$, $\mbox{Suffix} = m_{15} \| m_{23} \| m_{32}$,
and $H(M) = h_{41} = h$.

\section{Secret sharing schemes}

Based on a $(t+1, n)$ threshold scheme, many properties of secret
sharing schemes can be easily demonstrated. It has a simple access
structure and basis (Section 3.2). It is perfect and ideal (Section
3.3) and can be further implemented as proactive (Section 3.4) or
verifiable (Section 3.5). Also, the distribution of the shares and
recovery of the secret are through polynomial evaluation and
polynomial interpolation, respectively, which are easy to follow.

\subsection{A $(t+1, n)$ threshold scheme}

In 1979 Shamir \cite{SHA79} proposed a $(t+1, n)$ threshold scheme,
under which each of the $n$ participants $P_1, P_2, \ldots, P_n$
receives a share of the secret and any group of $t + 1$ or more
participants $(t \leq n -1)$ can recover the secret. Any group of
fewer than $t + 1$ participants cannot recover the secret. The
concept used by Shamir is based on Lagrange polynomial
interpolation. We generate a polynomial of degree at most $t$ over
$\mathbb{Z}_q$, where $q$ is a large prime number ($q > n \geq t +
1$). The coefficients, $a_t, \ldots, a_1 \in \mathbb{Z}_q$, are
generated randomly and $a_0 \in \mathbb{Z}_q$ is the secret.

\begin{equation}
P(x) = a_t x^t + a_{t-1} x^{ t-1} + \ldots + a_1 x^1 + a_0 \;
(\mbox{mod}\;q).
\end{equation}

The dealer arbitrarily chooses different $x_i \in \mathbb{Z}_q -
\{0\}, \; i=1, 2, \ldots, n$, and stores them in a public area. The
corresponding shares $P(x_i) (\mbox{mod}\;q)$ are then calculated
and distributed to the participants privately, so that each
participant gets a share of the secret. By the polynomial
interpolation given any $t + 1$ points the polynomial coefficients
can be recovered, hence the constant term $a_0$ which is the secret.
Note that we want the $n$ points to be all different to each other
and the coefficients must be from the field $\mathbb{Z}_q$ to make
sure we can recover the original polynomial. Here, we don't want to
give out the point $P(0)$, because $P(0)$ is the secret itself.

\subsection{Access structure}{\label{Sec:Access}}

Continuing with the construction above, it is reasonable to assume
that any number of greater than $t + 1$ participants can always
recover the secret. We call this property monotone. A group of
participants, which can recover the secret when they join together,
is called an authorized subset. In a $(t+1, n)$ threshold scheme,
any group of $t + 1$ or more participants forms an authorized
subset. On the other hand, any group of participants that cannot
recover the secret is called an unauthorized subset. An access
structure $\Gamma$ is a set of all authorized subsets.

Given any access structure $\Gamma$, $A \in \Gamma$ is called a
minimal authorized subset if $B \subsetneq A$ then $B \notin
\Gamma$. We use $\Gamma_0$, for the basis of $\Gamma$, to denote the
set of all minimal authorized subsets of $\Gamma$. In a $(t+1, n)$
threshold scheme, let $P$ be the set of the participants:

\begin{equation}
\Gamma = \{A\mid A \subseteq P \; \mbox{and} \; |A| \geq (t + 1)\},
\end{equation}
\begin{equation}
\Gamma_0 = \{A\mid A \subseteq P \; \mbox{and} \; |A| = (t+1)\}.
\end{equation}

In secret sharing, we first define the access structure. Then, we
realize the access structure by a secret sharing scheme. For
instance, Shamir's $(t+1, n)$ threshold scheme realizes the access
structure defined in Eq.4.

\subsection{Perfect and ideal scheme}

Shamir's scheme does not allow partial information to be given out
even up to $t$ participants joining together \cite{STI05}. A scheme
with such a property is called a perfect secret sharing scheme.
Based on information theory, the length of any share must be at
least as long as the secret itself in order to have perfect secrecy.
The argument for this is that up to $t$ participants have zero
information under the perfect sharing scheme, but when one extra
participant joins the group, the secret can be recovered. That means
any participant has his share at least as long as the secret. If the
shares and the secret come from the same domain, we call it an ideal
secret sharing scheme. In this case, the shares and the secret have
the same size.

\subsection{Proactive scheme}

In a secret sharing scheme, we need to consider the possibility that
an active adversary may find out all the shares in an authorized
subset to discover the secret eventually if he is allowed to have a
very long time to gather the necessary information. In order to
prevent this from happening, we refresh and redistribute new shares
to all the participants periodically. After finishing this phase,
the old shares are erased safely. The secret remains unchanged. By
doing so, the information gathered by the adversary between two
resets would be useless. In order to break the system an adversary
has to get enough information regarding the shares within any two
periodic resets.

Based on Shamir's scheme, Herzberg, Jarecki , Krawczyk, and Yung
\cite{HJKY95} derived a proactive scheme, which uses the following
method to renew the shares. In addition to $P(x)$ of Eq.3, the
dealer generates another polynomial $Q(x)$ of degree at most $t$
over $\mathbb{Z}_q$ without the constant term (i.e., $b_0 = 0$),

\begin{equation}
Q(x) = b_t x^t + b_{t-1} x^{t-1} + \ldots + b_1 x \;
(\mbox{mod}\;q),
\end{equation}

\noindent where $b_1, \ldots, b_t \in \mathbb{Z}_q$. Then add $P(x)$
and $Q(x)$ together to get the sum $R(x)$ as

\begin{equation}
R(x) = c_t x^t + c_{t-1} x^{t-1} + \ldots + c_1 x + a_0 \;
 (\mbox{mod}\;q),
\end{equation}

\noindent where $c_i = a_i + b_i$ (mod $q$)  for $i = 1, \ldots ,
t.$

The dealer then sends out new shares $R(1), R(2), \ldots,R(n)$ to
the $n$ participants to replace the old shares $P(0), P(1), \ldots,
P(n)$. It remains a $(t+1, n)$ threshold scheme with the same
original secret.

The above technique can be extended so that all the participants can
engage in the shares renewal process. This method can eliminate the
situation where all the work is done by the dealer, and the scheme
will be more secure.

\subsection{Verifiable scheme}

In reality, we need to consider the situation that the dealer or
some of the participants might be malicious. In this case, we need
to set up a verifiable secret sharing scheme so that the validity of
the shares can be verified. Here we discuss Feldman's scheme
\cite{FEL87} which is a simple verifiable secret sharing scheme
based on Shamir's scheme. Also see \cite{HL09} for another
reference.

The idea is to find a cyclic group $G$ of order $q$ where $q$ is a
prime. Since it is cyclic, a generator of $G$, say $g$, exists. As
other cryptographic protocols, we assume the parameters of $G$ are
carefully chosen so that the discrete logarithm problem is hard to
solve in $G$. Let $p, q$ be primes such that $q$ divides $(p-1), \;
g \in \mathbb{Z}_p^*$ of order $q$. The dealer generates a
polynomial $P(x)$ over $\mathbb{Z}_q$ of degree at most $t$ as shown
in Eq.3, and sends out $P(i)$ to participant $i$ as before. In
addition to this, he also broadcast in a public channel the
commitments: $g^{a_i} (\mbox{mod}\;p), \; i = 1, \ldots, n$. Each
participant $i$ will verify if the following equation is true.

\begin{equation}
g^{P(i)} = (g^{a_0})(g^{a_1})^i (g^{a_2})^{i^2} \ldots
       (g^{a_t})^{i^t} \; (\mbox{mod}\;p), i = 1, \ldots, n.
\end{equation}

Based on the homomorphic properties of the exponentiation, the above
condition will hold true if the dealer sends out consistent
information. Later, when the participants return their shares for
secret recovering, the dealer can also verify their shares by the
same method. Feldman's scheme is not perfect since partial
information about the secret, $g^{a_0}$, is leaked out. However we
assume it is difficult to get the secret $a_0$ from $g^{a_0}$ if the
discrete logarithm problem is hard to solve under $G$.

\section{Hash function based secret sharing scheme designs}

\subsection{Related work}

Zheng, Hardjono and Seberry \cite{ZHS94} discuss how to reuse shares
in a secret sharing scheme by using the universal hash function.
Chum and Zhang \cite{CZ10a,CZ10c} show how to apply hash functions
to Latin square based secret sharing schemes for improvements. In
this paper, we extend idea of herding and Nostradamus attacks
\cite{KK06} to any secret sharing scheme. We propose how to speed up
the process and hence make it practical. An outline of the
implementation is also suggested.

One direction for research in secret sharing schemes is to reduce
the size of the shares. One approach is to use a ramp scheme
\cite{IY06,KKFT09}. However, the limitation of a ramp scheme is that
it leaks partial information. If we want the scheme to be perfect as
aforementioned, the size of the share should be at least as long as
the size of the secret. It has been shown that there are no ideal
schemes for certain access structures. Please refer to
\cite{BL88,CSGV93,C97} for examples. That means at least one
participant needs to hold a share whose size is longer than the
secret. Here we want to set up an ideal scheme for any access
structure with the aid of a public area which is justified, because
of its relatively low cost to maintain. As long as no one can change
or destroy the public area it will work. To be explained later, the
public area does not help any authorized subset, with just one
participant missing, to recover the secret is easier than an
outsider if the length of the share is the same as that of the hash.
From the public area, we can only identify which group of
participants can be joined together to recover the secret. In
general, this is not a problem. In reality, should a secret need to
be accessed, we know who should be contacted. Our scheme is flexible
and fast because it makes use of the properties of the existing hash
functions.

\subsection{A simplified diamond structure}{\label{SSec:DiamStru}}

\begin{figure}
\centering
\includegraphics[height=6cm, width=8.5cm]{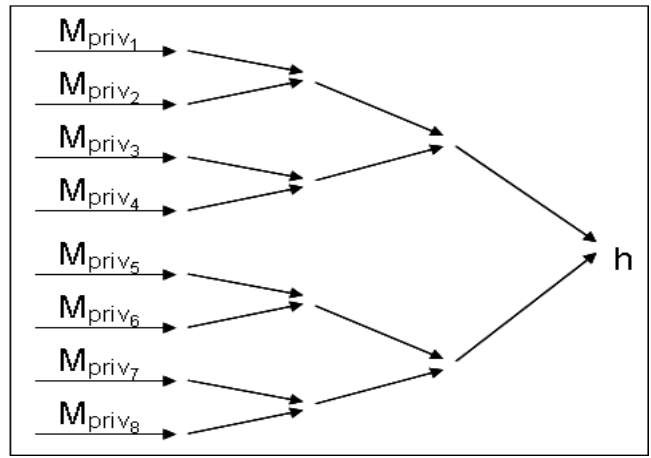}
\caption{Any authorized subset will herd to the final hash, i.e. the
secret.} \label{fig:DIAM3}
\end{figure}

In the proposed new scheme we set up one message $M_{priv}$ for one
authorized subset. After building a diamond structure, all the
$M_{priv}$'s will be herded to a final hash $h$, which is the
secret. That means any authorized subset can recover the secret by
their private shares and the corresponding public information (see
Fig.~\ref{fig:DIAM3}). More details are in the next section.

Based on the birthday attack the complexity of building a diamond
structure for our scheme is exponential, too expensive to implement.
We will show how to avoid such complexity and make the scheme
efficient and practical in next section.

\subsection{Newly proposed scheme}

\underline{\textbf{A. Setup}}

(a) We randomly generate a share of the same size as that of the
hash to each participant. Suppose there are $n$ participants, then
share $s_i$ will be assigned to participant $P_i$, $i=1, \ldots, n$.

(b) We determine all the minimal authorized subsets. Suppose we have
$A_1, \ldots, A_w$ minimal authorized subsets. Each participant
holds a share and combination of the shares of any one of these $w$
authorized subsets will form a private message $M_{priv}$. The
combination will be the concatenation of the shares in participant
sequence. For example, if an authorized subset consists of $P_1,
P_3$ and $P_4$, then $M_{priv} = s_1 || s_3 || s_4$.

(c) Calculate the hashes for the following
\begin{equation}
H(M_{priv_i}) = h_i,\; i = 1, \ldots, w.
\end{equation}

Let $h$ be the secret and of the same size of $h_i$. If we want the
secret to be random, we can set $h$ to one of the $h_i$. Or $h$ is a
pre-determined fixed secret. We continue to generate a control $c_i$
as follows (here $\oplus$ is bitwise exclusive OR):

\begin{equation}
c_i = h_i \oplus h, \; i = 1, \ldots, w.
\end{equation}

To summarize, after the setup process each participant $P_i$ gets a
random share $s_i, i = 1, \ldots, n$. Public information $c_i$,
where $i = 1, \ldots, w$, is generated. Control area $c_i$'s help to
herd all the intermediate hashes $h_i$'s to the final hash $h$. This
eliminates the complexity of building a diamond structure.\\

\underline{\textbf{B. Secret recovering}}

\begin{figure}
\centering
\includegraphics[height=2cm, width=8cm]{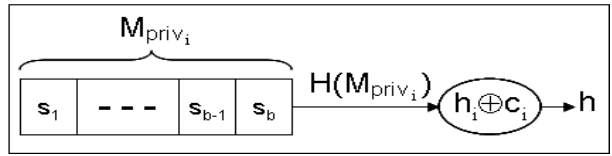}
\caption{Secret recovery by combination of private and public
information.} \label{fig:SecRecovery}
\end{figure}

Suppose authorized subset $A_i$ consists of participants $P_1,
\ldots, P_b$. Joining together they can recovery the secret as
follows, see Fig.~\ref{fig:SecRecovery}.

\begin{enumerate}[1)]
\item Get the public information $c_i$.
\item $H(s_1 || s_2 || \ldots || s_b) = h_i$, and $h_i
\oplus c_i = h$.
\end{enumerate}

\begin{figure}
\centering
\includegraphics[height=4cm, width=7cm]{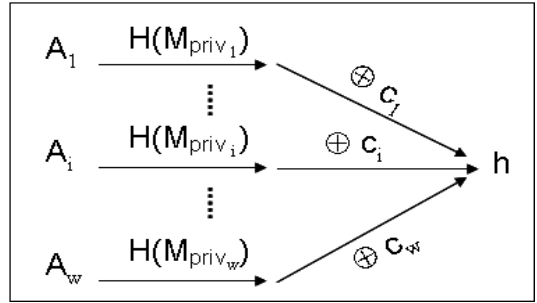}
\caption{Secret recovery for any authorized subset.}
\label{fig:SecretRG}
\end{figure}

This applies to any authorized subset, see Fig.~\ref{fig:SecretRG}.

\underline{\textbf{C. Performance}}

In the setup step the operations involved are generation of random
shares $s_1, \ldots, s_n$, calculation of hashes $h_i =
H(M_{priv_i})$, and generation of control area $c_i, i = 1, 2,
\ldots, w$. In the secret recovering step, assuming participants of
authorized subset $A_i$ join together, we just need to calculate the
hash (i.e., secret) by $h = H(M_{priv_i}) \oplus c_i$. All the
operations during the setup and secret recovering are efficient.
This makes the proposed scheme practical.\\

\underline{\textbf{D. Properties of the proposed scheme}}

\begin{enumerate}[a)]

\item Perfect: Based on randomness of a hash function, any participant cannot
figure out any information about the hash from his/her share.
Suppose a participant in a minimal authorized subset is missing, the
randomness property makes it impossible to recover his/her share
directly. Brute force is the only way to determine the share of the
missing participant. However, the rest of the participants cannot
rule out any possibility of the value of the share, as each guessed
value can be combined with their shares come up to a valid hash. So,
in the worst case, they need to try $2^{|s|}$ times. On the other
hand, an outsider needs to try $2^{|h|}$ times. If we choose the
size of the share $s$ as same as that of the hash $h$, any
authorized subset with just one participant missing does not have
any additional information to help them do better than any outsider.

\item Ideal: Each participant holds one share which has the same size of the
hash. The smaller the size of the shares $|s|$, the more efficient
the scheme would be. However, as discussed above, any authorized
subset with just one participant missing can recover the hash by
trying at most $2^{|s|}$ times. That means they can break the system
more easily than an outsider if $|s|$ is smaller than $|h|$. On the
other hand, it will not increase the security level by setting $|s|$
larger than $|h|$. By brute force, any outsider can try at most
$2^{|h|}$ times to recover the secret.

\item Fast setup and recovery of the secret: The calculation of
hash function is fast. No complicated or intensive computation, such
as polynomial evaluation/interpolation, is needed.

\item Application of minimal authorized subset: As we explained
earlier, we can speed up the whole process by considering the
minimal authorized subset only.

\item General access structure: As we shall see in the following
examples, this approach can be extended to any general access
structure.

\item Flexible: A hash function can handle any message of arbitrary length
so there is no limit to the number of participants. We can always
change to a new and better hash function should it become available.
For example, we use SHA-2 now, when SHA-3 is available we can switch
to it.

\item No special hardware or software is required: For example, no need to
handle a large number or find a large prime, etc.

\end{enumerate}

\subsection{Set up an ideal perfect $(t + 1, n)$ threshold scheme \label{Sec:Perfect}}

As we mentioned before, a $(t+1, n)$ threshold scheme has a simple
access structure. Based on the monotone property, we only need to
consider $N = C(n, t+1)$ minimal authorized subset only. Here,

\begin{equation}
N = C(n, t+1) = \frac{n!}{(t+1)!(n-t-1)!}.
\end{equation}

\noindent \underline{Example}: A (2, 3) threshold scheme\\

\begin{figure}
\centering
\includegraphics[height=4cm, width=7cm]{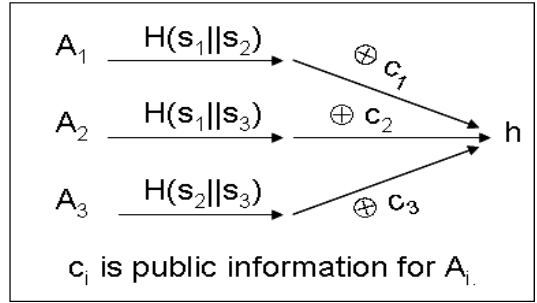}
\caption{A (2, 3) threshold scheme example.}
\label{fig:SimpleDiamond}
\end{figure}

Let $s_1, s_2$, and $s_3$ be shares of participants $P_1, P_2$, and
$P_3$, respectively. Then, the access structure consists of three
($N = 3$ by the Eq.11) minimal authorized subsets $A_1, A_2$ and
$A_3$. The controls $c_1, c_2, c_3$ will be stored in the public
area, see Fig.~\ref{fig:SimpleDiamond}.

\begin{enumerate}[a)]
\item $A_1:\; \{P_1, P_2\} \hspace{3.5em} s_1||s_2; \;c_1$
\item $A_2:\; \{P_1, P_3\} \hspace{3.5em} s_1||s_3; \;c_2$
\item $A_3:\; \{P_2, P_3\} \hspace{3.5em} s_2||s_3; \;c_3$
\end{enumerate}

\subsection{Set up an ideal perfect scheme for general access structure} {\label{Sec:SSSGeneral}}

Our herding hashes technique discussed above can be used to set up a
secret sharing scheme for any general access structure. Here, we
illustrate a hierarchical threshold scheme and a compartment scheme
as follows.

\subsubsection{Hierarchical threshold scheme}

The following is the conjunctive hierarchical scheme proposed by
Tassa \cite{TAS07}. Let $U$ be the set of $n$ participants. $U$ is
divided into $m$ levels:

\begin{equation} U = U_1 \cup U_2 \cup \ldots \cup U_m \; \mbox{and} \;U_i \cap U_j =
0, \forall i, j: 1 \leq i < j \leq m. \end{equation}

Instead of just assigning a threshold number $k$ as a regular secret
sharing scheme, a set of numbers $k=\{k_1, \ldots, k_m\}$ in a
strictly increasing order is set up: $0 < k_1 < k_2 < \ldots < k_m$.
Then, the $(k, n)$ hierarchical threshold access structure is:

\begin{equation} T = \{ V \subset U \mid\; | V \cap (U_1 \cup \ldots \cup U_i) |
\geq k_i, \forall i \in \{1,\ldots, m\} \}. \end{equation}

So if $V$ is an authorized subset, then:\\

\hspace{2em} the number of participants in $V$ at level 1 $ \geq k_1$ \\
AND \hspace{1em} the number of participants in V  at level 1, 2 $ \geq k_2$ \\
AND \hspace{1em} $\ldots \ldots \ldots \ldots \ldots \ldots \ldots \ldots$ \\
AND \hspace{1em} the number of participants in V at level $1,
\ldots, m \geq k_m$. \\

If we just require any one of the above conditions to be true at any
level, we can simply change AND to OR, then, we will get a
disjunctive hierarchical secret sharing scheme which is originally
proposed by Simmons \cite{SIM88}.\\

\noindent \underline{Example}: Conjunctive hierarchical secret
sharing scheme\\

\noindent Let $U = \{P_1, P_2, P_3, P_4, P_5, P_6\}$ be the set of
the participants. There are three levels, $U_1 = \{P_1, P_2\}$ for
level 1, $U_2 = \{P_3, P_4\}$ for level 2, $U_3 = \{P_5, P_6\}$ for
level 3, and $\{k_1, k_2, k_3\} = \{1, 2, 3\}$. Based on $\Gamma_0$,
the set of minimal authorized subsets, we have the following setup,
where $s_i$ is the corresponding share for $P_i$ and $c_i$'s are the
corresponding public information, $A_i$'s are authorized subsets.

\begin{enumerate}[a)]
\item $A_1:\;\; \{P_1, P_3, P_5\}$ \hspace{2em} $s_1 || s_3 || s_5; \;c_1$
\item $A_2:\;\; \{P_1, P_3, P_6\}$ \hspace{2em} $s_1 || s_3 || s_6; \;c_2$
\item $A_3:\;\; \{P_1, P_4, P_5\}$ \hspace{2em} $s_1 || s_4 || s_5; \;c_3$
\item $A_4:\;\; \{P_1, P_4, P_6\}$ \hspace{2em} $s_1 || s_4 || s_6; \;c_4$
\item $A_5:\;\; \{P_1, P_3, P_4\}$ \hspace{2em} $s_1 || s_3 || s_4; \;c_5$
\item $A_6:\;\; \{P_2, P_3, P_5\}$ \hspace{2em} $s_2 || s_3 || s_5; \;c_6$
\item $A_7:\;\; \{P_2, P_3, P_6\}$ \hspace{2em} $s_2 || s_3 || s_6; \;c_7$
\item $A_8:\;\; \{P_2, P_4, P_5\}$ \hspace{2em} $s_2 || s_4 || s_5; \;c_8$
\item $A_9:\;\; \{P_2, P_4, P_6\}$ \hspace{2em} $s_2 || s_4 || s_6; \;c_9$
\item $A_{10}:\; \{P_2, P_3, P_4\}$ \hspace{1.9em} $s_2 || s_3 || s_4; \;c_{10}$
\item $A_{11}:\; \{P_1, P_2, P_3\}$ \hspace{1.9em} $s_1 || s_2 || s_3; \;c_{11}$
\item $A_{12}:\; \{P_1, P_2, P_4\}$ \hspace{1.9em} $s_1 || s_2 || s_4; \;c_{12}$
\item $A_{13}:\; \{P_1, P_2, P_5\}$ \hspace{1.9em} $s_1 || s_2 || s_5; \;c_{13}$
\item $A_{14}:\; \{P_1, P_2, P_6\}$ \hspace{1.9em} $s_1 || s_2 || s_6; \;c_{14}$
\end{enumerate}

\subsubsection{Compartment scheme}

Compartment scheme \cite{SIM88} works as follows. Let $U$ be the set
of $n$ participants, and $U$ is divided into $m$ compartments: $ U =
U_1 \cup U_2 \cup \ldots \cup U_m$  and $U_i \cap U_j = 0 $ for all
$i, j: 1 \leq i < j \leq m .$

There is a threshold assigned to each group, say $t_1$ for $U_1$,
$t_2$ for $U_2$, etc. An authorized subset will:

\begin{enumerate}[a)]
\item contain at least $t_i$ participants in $U_i$ (an
individual threshold scheme for group $U_i$);
\item contain at least $t$ participants (an overall threshold
scheme).
\end{enumerate}

\noindent \underline{Example}: Compartment secret sharing scheme\\

\noindent Let $U = \{P_1, P_2, \ldots, P_6\}$ be the set of the
participants, three compartments $U_1 = \{P_1, P_2\}$, $U_2 = \{P_3,
P_4\}$ and $U_3 = \{P_5, P_6\}$. We want at least $1$ participant
from each compartment and $4$ participants overall. Once we
determine the $\Gamma_0$, the implementation will be
straightforward.

\begin{enumerate}[a)]
\item $A_1:\;\; \{P_1, P_2, P_3, P_5\}$ \hspace{2em} $s_1 || s_2 || s_3 || s_5; \;c_1$
\item $A_2:\;\; \{P_1, P_2, P_3, P_6\}$ \hspace{2em} $s_1 || s_2 || s_3 || s_6; \;c_2$
\item $A_3:\;\; \{P_1, P_2, P_4, P_5\}$ \hspace{2em} $s_1 || s_2 || s_4 || s_5; \;c_3$
\item $A_4:\;\; \{P_1, P_2, P_4, P_6\}$ \hspace{2em} $s_1 || s_2 || s_4 || s_6; \;c_4$
\item $A_5:\;\; \{P_1, P_3, P_4, P_5\}$ \hspace{2em} $s_1 || s_3 || s_4 || s_5; \;c_5$
\item $A_6:\;\; \{P_1, P_3, P_4, P_6\}$ \hspace{2em} $s_1 || s_3 || s_4 || s_6; \;c_6$
\item $A_7:\;\; \{P_2, P_3, P_4, P_5\}$ \hspace{2em} $s_2 || s_3 || s_4 || s_5; \;c_7$
\item $A_8:\;\; \{P_2, P_3, P_4, P_6\}$ \hspace{2em} $s_2 || s_3 || s_4 || s_6; \;c_8$
\item $A_9:\;\; \{P_1, P_3, P_5, P_6\}$ \hspace{2em} $s_1 || s_3 || s_5 || s_6; \;c_9$
\item $A_{10}:\; \{P_1, P_4, P_5, P_6\}$ \hspace{1.9em} $s_1 || s_4 || s_5 || s_6; \;c_{10}$
\item $A_{11}:\; \{P_2, P_3, P_5, P_6\}$ \hspace{1.9em} $s_2 || s_3 || s_5 || s_6; \;c_{11}$
\item $A_{12}:\; \{P_2, P_4, P_5, P_6\}$ \hspace{1.9em} $s_2 || s_4 || s_5 || s_6; \;c_{12}$
\end{enumerate}

\subsection{Set up a verifiable scheme for general access structure \label{Sec:Verifiable}}

Let $f, g$ be cryptographic hash functions. The dealer generates
shares $s_1, s_2, \ldots,$ and distributes each share to each
participant and then publishes the hashes (by hash function $g$) of
each share as commitments: $g_1, g_2, \ldots$. Participant $i$
verifies his or her share by checking if $g(m_i) = g_i$ holds. If
all participants confirm that taking his or her share as input to
the hash function $g$, he or she gets the hash value equal to one of
the commitments published by the dealer, we conclude the dealer
sends out consistent shares. Likewise, when the participants return
their shares, the dealer can verify in the same way.

Hash function $g$ is used to make the scheme verifiable. Hash
function $f$ is used as $H$ in 4.3 for the scheme. Partial
information was given out here, however, if $g$ is preimage
resistant, it would be infeasible to find the original share $s_i$
from $g_i$. Participant $i$ can fool the party if he or she can find
$s_i'$ such that $g(s_i) = g(s_i') = g_i$. However, this is also
extremely difficult to achieve if $g$ is second preimage resistant.

\subsection{Set up a proactive scheme}

We pick up any authorized subset to recover the secret $h$, then
repeat the process to generate and re-distribute new shares $s_1',
s_2', \ldots$. Based on the secret $h$ and the newly generated
shares $s_1', s_2', \ldots $, we determine and update the new public
control information $c_1', c_2', \ldots $. Finally we delete the
secret $h$. So shares are refreshed and the secret remains
unchanged.

\section{Implementation plan}

Suppose there are $n$ participants $P_1, \ldots, P_n$ and $w$
minimal authorized subsets $A_1, \ldots, A_w$ for a given access
structure. Let $H$ be the hash function for the implementation. The
secret stores in a variable $h$, which has the same size
as the output hash of $H$. \\

\noindent (a) If the secret is fixed, input and store it in $h$.
Otherwise skip this step.

\noindent (b) FOR $i = 1, \ldots, n$

\hspace{30pt} Generate randomly $s_i$ for $P_i$

ENDFOR \\

\noindent (c) FOR $i = 1, \ldots, w$

\hspace{30pt} Construct $M_{priv_i}$ based on shares of participants
in $A_i$ in participant sequence

\hspace{30pt} $h_i = H(M_{priv_i})$

\hspace{30pt} If $i = 1$ and $h$ is empty, then $h = h_1$ /* If no
input secret, set the secret to the first randomly generated
intermediate hash $h_1$. */

\hspace{30pt} $c_i = h_i \oplus h$

\hspace{30pt} $K_i = $ concatenation of the ordered indices of
participants in $A_i$

\hspace{30pt} Write $c_i$ in public area based on key $K_i$

ENDFOR \\

\noindent (d) FOR $i = 1, \ldots, n$

\hspace{30pt} Send $s_i$ to $P_i$ privately

ENDFOR \\

\noindent (e) Delete all $s_i$ (shares) and the $h$ (secret).\\

\begin{figure}
\centering
\includegraphics[height=4cm, width=8cm]{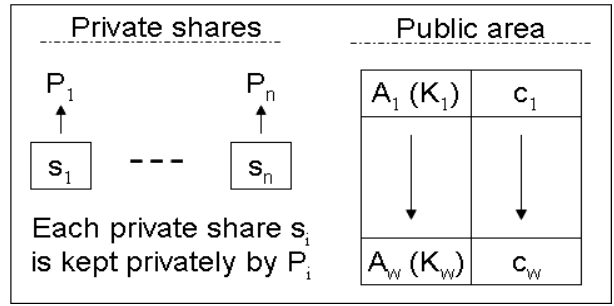}
\caption{Shares for participants and public area.} \label{fig:Setup}
\end{figure}

\noindent After the implementation:

\begin{enumerate}[1)]
\item We create the following (see Fig.~\ref{fig:Setup})
\begin{enumerate}[i)]
\item private shares for participants: $s_1, \ldots, s_n$.
\item public information $c_1, \ldots,
c_w$.
\end{enumerate}
\item Any authorized subset $A_i$ can form key $K_i$ to get the
corresponding $c_i$ to recover the secret (see 4.3B).
\end{enumerate}

\section{Conclusion}

This paper shows how to design various secret sharing schemes based
on cryptographic hash functions so that any general access structure
can be realized as perfect and ideal. The implementation is simple
and efficient as we make use of the existing hash functions. The
share distribution and secret recovery can be done quickly due to
fast calculation of hash functions. We can further implement these
schemes as proactive and, or verifiable if required.

\subsubsection*{Acknowledgments.} Authors would like to thank
Michelle Baker for her comments to the initial draft, Vincent Falco
for proofreading the manuscript, reviewers for their valuable
suggestions. The second author's work was supported in part by
PSC-CUNY Research Award: 63574-0041.

\bibliographystyle{unsrt}
\bibliography{Chi_Sean}

\begin{thebibliography}{10}

\bibitem{KL07}
J.~Katz and Y.~Lindell.
\newblock {\em Introduction to Modern Cryptography}.
\newblock Chapman and Hall/CRC, 2007.

\bibitem{AAEM09}
D.~Agrawal, A.~Abbadi, F.~Emekci, and A.~Metwally.
\newblock Database management as a service: challenges and opportunities.
\newblock In {\em IEEE ICDE}, 2009.

\bibitem{YAO82}
A.C. Yao.
\newblock Protocols for secure computations (extended abstract).
\newblock In {\em the 21st Annual IEEE Symposium on the Foundations of Computer
  Science}, pages 160--164, 1982.

\bibitem{CDN09}
R.~Cramer, I.~Damg{\aa}rd, and J.B. Nielsen.
\newblock {Multiparty Computation, an Introduction}.
\newblock Lecture Notes. Available http://www.brics.dk/$\sim$jbn/smc.pdf, 2009.

\bibitem{STI05}
D.~Stinson.
\newblock {\em Cryptography, Theory and Practice}.
\newblock Chapman and Hall/CRC, 3rd edition, 2005.

\bibitem{TW06}
W.~Trappe and L.~Washington.
\newblock {\em Introduction to Cryptography with Coding Theory}.
\newblock Prentice Hall, 2nd edition, 2006.

\bibitem{DAM89}
I.~Damg{\aa}rd.
\newblock A design principle for hash functions.
\newblock In {\em Proc. of CRYPTO 1989}, volume 435 of {\em LNCS}.

\bibitem{MER79}
R.~C. Merkle.
\newblock {\em Secrecy, Authentication, and Public Key Systems}.
\newblock PhD thesis, Stanford University, 1979.

\bibitem{JOUX04}
A.~Joux.
\newblock {Multicollisions in iterated hash functions. Application to cascaded
  construction}.
\newblock In {\em Proc. of CRYPTO 2004}, volume 3152 of {\em LNCS}, pages
  306--316.

\bibitem{KK06}
J.~Kelsey and T.~Kohno.
\newblock Herding hash functions and the {N}ostradamus attack.
\newblock Cryptology ePrint Archive, Report 2005/281, 2005.

\bibitem{SLW07}
M.~Stevens, A.K. Lenstra, and B.~Weger.
\newblock {Predicting the winner of the 2008 US presidential elections using a
  Sony PlayStation 3}.
\newblock Available online http://www.win.tue.nl/hashclash/Nostradamus,
  November 2007.

\bibitem{SHA79}
A.~Shamir.
\newblock How to share a secret.
\newblock {\em Communications of the ACM}, 22(11):612--613, 1979.

\bibitem{HJKY95}
A.~Herzberg, S.~Jarecki, H.~Krawczyk, and M.~Yung.
\newblock Proactive secret sharing.
\newblock In {\em Proc. of CRYPTO 1995}, volume 963 of {\em LNCS}, 1995.

\bibitem{FEL87}
P.~Feldman.
\newblock A practical scheme for non-interactive verifiable secret sharing.
\newblock In {\em Proc. of the 28th IEEE Symposium on the Foundations of
  Computer Science}, pages 427--437, 1987.

\bibitem{HL09}
L.~Harn and C.~Lin.
\newblock Detection and identification of cheaters in $(t, n)$ secret sharing
  scheme.
\newblock {\em Designs, Codes and Cryptography}, 52(1):15--24, 2009.

\bibitem{ZHS94}
Y.~Zheng, T.~Hardjono, and J.~Seberry.
\newblock Reusing shares in secret sharing schemes.
\newblock {\em Computer Journal}, 37(3):199--205, 1994.

\bibitem{CZ10a}
C.~Chum and X.~Zhang.
\newblock {Applying hash functions in the Latin square based secret sharing
  schemes}.
\newblock In {\em Proc. of The 2010 International Conference on Security and
  Management (SAM'10)}, pages 197--203, 2010.

\bibitem{CZ10c}
C.~Chum and X.~Zhang.
\newblock {The Latin squares and the secret sharing schemes}.
\newblock {\em J. of Groups - Complexity - Cryptology}, 2:175--202, 2010.

\bibitem{IY06}
M.~Iwamoto and H.~Yamamoto.
\newblock Strongly secure ramp secret sharing schemes for general access
  structures.
\newblock {\em Journal Information Processing Letters}, 97(2), 2006.

\bibitem{KKFT09}
J.~Kurihara, S.~Kiyomoto, K.~Fukushima, and T.~Tanaka.
\newblock A fast (k, l, n)-threshold ramp secret sharing scheme.
\newblock {\em IEICE Transactions on Fundamentals of Electronics,
  Communications and Computer Sciences}, E92.A(8):1808--1821, 2009.

\bibitem{BL88}
J.~Benaloh and J.~Leichter.
\newblock Generalized secret sharing and monotone functions.
\newblock In {\em CRYPTO 1988}, volume 403 of {\em LNCS}, pages 27--35, 1990.

\bibitem{CSGV93}
R.M. Capocelli, A.~De Santis, L.~Gargano, and U.~Vaccaro.
\newblock On the size of shares for secret sharing schemes.
\newblock {\em Journal of Cryptology}, 6(3):157--167, 1993.

\bibitem{C97}
L.~Csirmaz.
\newblock The size of a share must be large.
\newblock {\em Journal of Cryptology}, 10(4):223--231, 1997.

\bibitem{TAS07}
T.~Tassa.
\newblock Hierarchical threshold secret sharing.
\newblock {\em Journal of Cryptology}, 20(11):237--264, 2007.

\bibitem{SIM88}
G.J. Simmons.
\newblock How to (really) share a secret.
\newblock In {\em CRYPTO1988}, volume 403 of {\em LNCS}, pages 390--448, 1990.

\end{thebibliography}

\end{document}